\documentstyle[12pt]{article}
\def\1{{\chi}}

\begin{document}
\title {{Sequential product on standard effect algebra ${\cal E} (H)$}\thanks{This project is supported by Natural Science
Found of China (10771191 and 10471124).}}
\author {Shen Jun$^{1,2}$, Wu Junde$^{1}$\date{}\thanks{E-mail: wjd@zju.edu.cn}}
\maketitle
$^1${\small\it Department of Mathematics, Zhejiang
University, Hangzhou 310027, P. R. China}

$^2${\small\it Department of Mathematics, Anhui Normal University,
Wuhu 241003, P. R. China}

\begin{abstract} {\noindent A quantum effect is an operator $A$ on a complex Hilbert space $H$ that satisfies $0\leq A\leq
I$, ${\cal E} (H)$ is the set of all quantum effects on $H$. In
2001, Professor Gudder and Nagy studied the sequential product
$A\circ B=A^{\frac{1}{2}}BA^{\frac{1}{2}}$ of $A, B\in {\cal E}
(H)$. In 2005, Professor Gudder asked: Is $A\circ
B=A^{\frac{1}{2}}BA^{\frac{1}{2}}$ the only sequential product on
${\cal E} (H)$? Recently, Liu and Wu presented an example to show
that the answer is negative. In this paper, firstly, we characterize
some algebraic properties of the abstract sequential product on
${\cal E} (H)$; secondly, we present a general method for
constructing sequential products on ${\cal E} (H)$; finally, we
study some properties of the sequential products constructed by the
method.}
\end{abstract}


\vskip0.4in

{\bf 1. Introduction}

\vskip0.2in

Sequential effect algebra is an important model for studying the
quantum measurement theory ([1-7]). A sequential effect algebra is
an effect algebra which has a sequential product operation. Firstly,
we recall some elementary notations and results.

An {\it effect algebra} is a system $(E,0,1, \oplus)$, where 0 and 1
are distinct elements of $E$ and $\oplus$ is a partial binary
operation on $E$ satisfying that [8]:

(EA1) If $a\oplus b$ is defined, then $b\oplus a$ is defined and
$b\oplus a=a\oplus b$.

(EA2) If $a\oplus (b\oplus c)$ is defined, then $(a\oplus b)\oplus
c$ is defined and $$(a\oplus b)\oplus c=a\oplus (b\oplus c).$$

(EA3) For each $a\in E$, there exists a unique element $b\in E$ such
that $a\oplus b=1$.

(EA4) If $a\oplus 1$ is defined, then $a=0$.

\vskip 0.1 in

In an effect algebra $(E,0,1, \oplus)$, if $a\oplus b$ is defined,
we write $a\bot b$. For each $a\in (E,0,1, \oplus)$, it follows from
(EA3) that there exists a unique element $b\in E$ such that $a\oplus
b=1$, we denote $b$ by $a'$. Let $a, b\in (E,0,1, \oplus)$, if there
exists a $c\in E$ such that $a\bot c$ and $a\oplus c=b$, then we say
that $a\leq b$. It follows from [8] that $\leq $ is a partial order
of $(E,0,1, \oplus)$ and satisfies that for each $a\in E$, $0\leq
a\leq 1$, $a\bot b$ if and only if $a\leq b'$.

Let $(E,0,1, \oplus , \circ)$ be an effect algebra and $a\in E$. If
$a\wedge a'=0$, then $a$ is said to be a {\it sharp element} of $E$.
We denote $E_S$ the set of all sharp elements of $E$ ([9-10]).

\vskip 0.1 in

A {\it sequential effect algebra} is an effect algebra $(E,0,1,
\oplus)$ with another binary operation $\circ $ defined on it
satisfying [2]:

(SEA1) The map $b\mapsto a\circ b$ is additive for each $a\in E$,
that is, if $b\bot c$, then $a\circ b\bot a\circ c$ and $a\circ
(b\oplus c)=a\circ b\oplus a\circ c$.

(SEA2) $1\circ a=a$ for each $a\in E$.

(SEA3) If $a\circ b=0$, then $a\circ b=b\circ a$.

(SEA4) If $a\circ b=b\circ a$, then $a\circ b'=b'\circ a$ and
$a\circ (b\circ c)=(a\circ b)\circ c$ for each $c\in E$.

(SEA5) If $c\circ a=a\circ c$ and $c\circ b=b\circ c$, then
$c\circ(a\circ b)=(a\circ b)\circ c$ and $c\circ(a\oplus b)=(a\oplus
b)\circ c$ whenever $a\bot b$.

\vskip 0.1 in

If $(E,0,1, \oplus, \circ)$ is a sequential effect algebra, then the
operation $\circ$ is said to be a {\it sequential product} on
$(E,0,1, \oplus)$. If $a, b\in (E,0,1, \oplus, \circ)$ and $a\circ
b=b\circ a$, then $a$ and $b$ is said to be {\it sequentially
independent} and is denoted by $a|b$ ([1-2]).

\vskip 0.1 in

Let $H$ be a complex Hilbert space, ${\cal B}(H)$ be the set of all
bounded linear operators on $H$, ${\cal P}(H)$ be the set of all
projections on $H$, ${\cal E} (H)$ be the set of all self-adjoint
operators on $H$ satisfying that $0\leq A\leq I$. For $A,B\in {\cal
E} (H)$, we say that $A\oplus B$ is defined if $A+B\in {\cal E}
(H)$, in this case, we define $A\oplus B=A+B$. It is easy to see
that $({\cal E} (H),0,I,\oplus)$ is an effect algebra, we call it
{\it{standard effect algebra}} ([8]). Each element $A$ in ${\cal E}
(H)$ is said to be a {\it quantum effect}, the set ${\cal E} (H)_S$
of all sharp elements of $({\cal E} (H),0,I,\oplus)$ is just ${\cal
P}(H)$ ([2, 9]).

\vskip 0.1 in

Let $A\in {\cal B}(H)$, we denote $Ker(A)=\{x\in H\mid Ax=0\}$,
$Ran(A)=\{Ax\mid x\in H\}$, $P_{Ker(A)}$ denotes the projection onto
$Ker(A)$. Let $x\in H$ be a unit vector, $P_x$ denotes the
projection onto the one-dimensional subspace spanned by $x$.

\vskip 0.1 in

In 2001 and 2002, Professor Gudder, Nagy and Greechie showed that
for any two quantum effects $A$ and $B$, if we define $A\circ
B=A^{\frac{1}{2}}BA^{\frac{1}{2}}$, then the operation $\circ$ is a
sequential product on the standard effect algebra $({\cal E}
(H),0,I,\oplus)$, moreover, they studied some properties of this
special sequential product on $({\cal E} (H),0,I,\oplus)$ ([1,2]).

In 2005, Professor Gudder asked ([4]): Is $A\circ
B=A^{\frac{1}{2}}BA^{\frac{1}{2}}$ the only sequential product on
standard effect algebra $({\cal E} (H),0,I,\oplus)$?

In 2009, Liu and Wu constructed a new sequential product on $({\cal
E} (H),0,I,\oplus)$, thus answered Gudder's problem negatively
([7]). This new sequential product on $({\cal E} (H),0,I,\oplus)$
motivated us to study the following topics: (1) Characterize the
algebraic properties of abstract sequential product on $({\cal E}
(H),0,I,\oplus)$. (2) Present a general method for constructing
sequential product on $({\cal E} (H),0,I,\oplus)$. (3) Characterize
some elementary properties of the sequential product constructed by
the method. Our results generalize many conclusions in [1,3,7,14].

\vskip 0.2 in

{\bf 2. Abstract sequential product on $({\cal E} (H),0,I,\oplus)$}

 \vskip 0.2 in

In this section, we study some elementary properties of the abstract
sequential product on the standard effect algebra $({\cal E}
(H),0,I,\oplus)$.

{\bf Lemma 2.1 ([2])}. Let $(E,0,1, \oplus, \circ)$ be a sequential
effect algebra, $a\in E$. Then the following conditions are all
equivalent:

(1) $a\in E_S$;

(2) $a\circ a'=0$;

(3) $a\circ a=a$.

\vskip 0.1 in

{\bf Lemma 2.2 ([2])}. Let $(E,0,1, \oplus, \circ)$ be a sequential
effect algebra, $a\in E$, $b\in E_S$. Then the following conditions
are all equivalent:

(1) $a\leq b$;

(2) $a\circ b=b\circ a=a$.

\vskip 0.1 in

{\bf Lemma 2.3 ([2, 8])}. Let $(E,0,1, \oplus, \circ)$ be a
sequential effect algebra, $a,b,c\in E$.

(1) If $a\perp b$, $a\perp c$ and $a\oplus b=a\oplus c$, then $b=c$.

(2) $a\circ b\leq a$.

(3) If $a\leq b$, then $c\circ a\leq c\circ b$.

\vskip 0.1 in

{\bf Lemma 2.4 ([7])}. Let $\circ$ be a sequential product on the
standard effect algebra $({\cal E} (H),0,I,\oplus)$. Then for any
$A,B\in {\cal E} (H)$ and real number $t$, $0\leq t\leq 1$, we have
$(tA)\circ B=A\circ (tB)=t(A\circ B)$.

\vskip 0.1 in

{\bf Lemma 2.5 ([1])}. Let $A,B,C\in {\cal B} (H)$ and $A,B,C$ be
self-adjoint operators. If for every unit vector $x\in H$, $\langle
Cx,x\rangle =\langle Ax,x\rangle \langle Bx,x\rangle $, then $A=tI$
or $B=tI$ for some real number $t$.

\vskip 0.1 in

{\bf Lemma 2.6 ([11])}. Let $A\in {\cal B}(H)$ have the following
operator matrix form $$A=\left(
      \begin{array}{cc}
        A_{11} & A_{12} \\
        A_{21} & A_{22} \\
      \end{array}
    \right)
$$ with respect to the space decomposition $H=H_1\oplus H_2$. Then $A\geq 0$ iff

(1) $A_{ii}\in {\cal B} (H_i)$ and $A_{ii}\geq 0$, $i=1,2$;

(2) $A_{21}=A_{12}^*$;

(3) there exists a linear operator $D$ from $H_2$ into $H_1$ such
that $||D||\leq 1$ and
$A_{12}=A_{11}^{\frac{1}{2}}DA_{22}^{\frac{1}{2}}$.

\vskip 0.1 in

\vskip 0.1 in

{\bf Theorem 2.1.} Let $\circ$ be a sequential product on $({\cal E}
(H),0,I,\oplus)$, $B\in {\cal E}(H)$, $E\in {\cal P}(H)$. Then
$E\circ B=EBE$.

{\bf Proof.} For $A\in {\cal E}(H)$, let $\Phi_A:{\cal
E}(H)\longrightarrow {\cal E}(H)$ be defined by $\Phi_A(C)=A\circ C$
for each $C\in {\cal E}(H)$. It follows from Lemma 2.4 and (SEA1)
that $\Phi_A$ is affine on the convex set ${\cal E}(H)$. Note that
${\cal E}(H)$ generates algebraically the vector space ${\cal
B}(H)$, so $\Phi_A$ has a unique linear extension to ${\cal B}(H)$,
which we also denote by $\Phi_A$. Then $\Phi_A$ is a positive linear
operator on ${\cal B}(H)$ and $\Phi_A(I)=A$. Thus $\Phi_A$ is
continuous.

Note that $E\in {\cal P}(H)={\cal E} (H)_S$, it follows from Lemma
2.1 that $E\circ (I-E)=0$ and so $\Phi_E(I-E)=0$. By composing
$\Phi_E$ with all states on ${\cal B}(H)$ and using Schwarz's
inequality, we conclude that $\Phi_E(B)=\Phi_E(EBE)$. Since $EBE\in
{\cal E}(H)$, $E\in {\cal E} (H)_S$ and $EBE\leq E$, by Lemma 2.2 we
have $E\circ (EBE)=EBE$. Thus $E\circ B=\Phi_E(B)=\Phi_E(EBE)=E\circ
(EBE)=EBE$.

\vskip 0.1 in

{\bf Theorem 2.2.} Let $\circ$ be a sequential product on $({\cal E}
(H),0,I,\oplus)$, $A, B\in {\cal E}(H)$ and $AB=BA$. Then $A\circ
B=B\circ A=AB$.

{\bf Proof.} We use the notations as in the proof of Theorem 2.1.

Suppose $E\in {\cal P}(H)$ and $E\in \{A\}'$, i.e., $EA=AE$. Note
that $EAE,(I-E)A(I-E)\in {\cal E}(H)$, $EAE\leq E$ and
$(I-E)A(I-E)\leq I-E$, by Lemma 2.2, it follows that $EAE|E$ and
$(I-E)A(I-E)|(I-E)$. Since $A=EAE+(I-E)A(I-E)$, by (SEA4) and (SEA5)
we have $A|E$. By Theorem 2.1 we conclude that $A\circ E=E\circ
A=EAE=AE$. Thus, $\Phi_A(E)=AE$. Since $\Phi_A$ is a continuous
linear operator and $\{A\}'$ is a von Neumann algebra, we conclude
that $\Phi_A(B)=AB$. That is, $A\circ B=AB$. Similarly, we have
$B\circ A=BA$. Thus $A\circ B=B\circ A=AB$.

\vskip 0.1 in

{\bf Theorem 2.3.}  Let $\circ$ be a sequential product on $({\cal
E} (H),0,I,\oplus)$, $A, B\in {\cal E}(H)$. Then the following
conditions are all equivalent:

(1) $AB=BA=B$;

(2) $A\circ B\geq B$;

(3) $A\circ B=B$;

(4) $B\circ A=B$;

(5) $B\leq P_{Ker(I-A)}$;

(6) $B\leq A^n$ for each positive integer $n$.

{\bf Proof.} (1)$\Rightarrow$(3) and (1)$\Rightarrow$(4): By Theorem
2.2.

(3)$\Rightarrow$(2) is obvious.

(4)$\Rightarrow$(3): By Theorem 2.2, $B\circ A=B=B\circ I$. Thus, it
follows from Lemma 2.3 that $B\circ (I-A)=0$. By (SEA3), $B|(I-A)$.
By (SEA4), $B|A$. So $A\circ B=B\circ A=B$.

(2)$\Rightarrow$(6): By using Theorem 2.2 and Lemma 2.3 repeatedly,
we have:

$B\leq A\circ B\leq A\circ I=A$;

$A\circ B\leq A\circ(A\circ B)\leq A\circ A=A^2$;

$A\circ(A\circ B)\leq A\circ\Big{(}A\circ(A\circ B)\Big{)}\leq
A\circ A^2=A^3$;

$\vdots$

$A\circ\cdots\circ(A\circ B)\leq
A\circ\Big{(}A\circ\cdots\circ(A\circ B)\Big{)}\leq A\circ
A^{n-1}=A^n$.

The above showed that $B\leq A^n$ for each positive integer $n$.

(6)$\Rightarrow$(5): Let $\chi_{\{1\}}$ be the characteristic
function of $\{1\}$. Note that $0\leq A\leq I$, it is easy to know
that $\{A^n\}$ converges to $\chi_{\{1\}}(A)=P_{Ker(I-A)}$ in the
strong operator topology. Thus $B\leq P_{Ker(I-A)}$.

(5)$\Rightarrow$(1): Since $0\leq B\leq P_{Ker(I-A)}$, we have
$Ker(P_{Ker(I-A)})\subseteq Ker(B)$. So $Ran(B)\subseteq
Ran(P_{Ker(I-A)})=Ker(I-A)$. Thus $(I-A)B=0$. That is, $AB=B$.
Taking adjoint, we get $AB=BA=B$.

\vskip 0.1 in

{\bf Theorem 2.4.} Let $\circ$ be a sequential product on $({\cal E}
(H),0,I,\oplus)$, $A, B\in {\cal E}(H)$. Then the following
conditions are all equivalent:

(1) $C\circ(A\circ B)=(C\circ A)\circ B$ for every $C\in{\cal
E}(H)$;

(2) $\langle (A\circ B)x,x\rangle =\langle Ax,x\rangle \langle
Bx,x\rangle $ for every $x\in H$ with $\|x\|=1$;

(3) $A=tI$ or $B=tI$ for some real number $0\leq t\leq 1$.

{\bf Proof.} By Lemma 2.5, we conclude that (2)$\Rightarrow$(3). By
Theorem 2.2 and Lemma 2.4, (3)$\Rightarrow$(1) is trivial.

(1)$\Rightarrow$(2): If (1) hold, then $P_x\circ(A\circ B)=(P_x\circ
A)\circ B$ for every $x\in H$ with $\|x\|=1$. By Theorem 2.1,
$P_x\circ(A\circ B)=P_x(A\circ B)P_x=\langle (A\circ B)x,x\rangle
P_x$. By Theorem 2.1 and Lemma 2.4, $(P_x\circ A)\circ
B=(P_xAP_x)\circ B=(\langle Ax,x\rangle P_x)\circ B=\langle
Ax,x\rangle (P_x\circ B)=\langle Ax,x\rangle P_xBP_x=\langle
Ax,x\rangle \langle Bx,x\rangle P_x$. Thus (2) hold.

\vskip 0.1 in

{\bf Theorem 2.5.} Let $\circ$ be a sequential product on $({\cal E}
(H),0,I,\oplus)$, $B\in {\cal E}(H)$, $E\in {\cal P}(H)$. Then the
following conditions are all equivalent:

(1) $E\circ B\leq B$;

(2) $EB=BE$;

(3) $E\circ B=B\circ E$.

{\bf Proof.} (2)$\Rightarrow$(3): By Theorem 2.2.

(3)$\Rightarrow$(1): By Lemma 2.3.

(1)$\Rightarrow$(2): Since $E\in {\cal P}(H)$, by Theorem 2.1,
$E\circ B=EBE$. Thus, $B-EBE\geq 0$. Note that $$B-EBE=\left(
      \begin{array}{cc}
        0 & EB(I-E) \\
        (I-E)BE & (I-E)B(I-E) \\
      \end{array}
    \right)
$$ with respect to the space decomposition $H=Ran(E)\oplus Ker(E)$, so by Lemma 2.6 we have $EB(I-E)=(I-E)BE=0$.
Thus $B=EBE+(I-E)B(I-E)$. So $EB=BE$.

\vskip 0.1 in

{\bf Theorem 2.6.} Let $\circ$ be a sequential product on $({\cal E}
(H),0,I,\oplus)$, $A, B, C\in {\cal E}(H)$. If $A$ is invertible,
then the following conditions are all equivalent:

(1) $B\leq C$;

(2) $A\circ B\leq A\circ C$.

{\bf Proof.} (1)$\Rightarrow$(2): By Lemma 2.3.

(2)$\Rightarrow$(1): It is easy to see that
$\|A^{-1}\|^{-1}A^{-1}\in {\cal E}(H)$.

By Lemma 2.3, $(\|A^{-1}\|^{-1}A^{-1})\circ(A\circ B)\leq
(\|A^{-1}\|^{-1}A^{-1})\circ(A\circ C)$.

By Theorem 2.2, $(\|A^{-1}\|^{-1}A^{-1})|A$ and
$(\|A^{-1}\|^{-1}A^{-1})\circ A=\|A^{-1}\|^{-1}I$.

By (SEA4) and Theorem 2.2, we have

$(\|A^{-1}\|^{-1}A^{-1})\circ(A\circ
B)=\Big{(}(\|A^{-1}\|^{-1}A^{-1})\circ A\Big{)}\circ
B=(\|A^{-1}\|^{-1}I)\circ B=\|A^{-1}\|^{-1}B$,

$(\|A^{-1}\|^{-1}A^{-1})\circ(A\circ
C)=\Big{(}(\|A^{-1}\|^{-1}A^{-1})\circ A\Big{)}\circ
C=(\|A^{-1}\|^{-1}I)\circ C=\|A^{-1}\|^{-1}C$.

So $B\leq C$.

\vskip 0.1 in

{\bf Corollary 2.1.}  Let $\circ$ be a sequential product on $({\cal
E}(H),0,I,\oplus)$, $A, B, C\in {\cal E}(H)$. If $A$ is invertible,
then the following conditions are all equivalent:

(1) $B=C$;

(2) $A\circ B=A\circ C$.

\vskip 0.2 in

{\bf 3. General method for constructing sequential products on
${\cal E}(H)$}

\vskip 0.2 in

In the sequel, unless specified, suppose $H$ be a finite dimensional
complex Hilbert space, ${\mathbf{C}}$ be the set of complex numbers,
${\mathbf{R}}$ be the set of real numbers, for each $A\in{\cal
E}(H)$, $sp(A)$ be the spectra of $A$ and $B(sp(A))$ be the set of
all bounded complex Borel functions on $sp(A)$.

Let $A,B\in {\cal B}(H)$, if there exists a complex constant $\xi$
such that $|\xi|=1$ and $A=\xi B$, then we denote $A\approx B$.

\vskip 0.1 in

In [7], Liu and Wu showed that if we define $A\circ
B=A^{\frac{1}{2}}f_i(A)B f_{-i}(A)A^{\frac{1}{2}}$ for $A, B\in
{\cal E}(H)$, where $f_z(t)=\exp z(\ln t)$ if $t\in (0, 1]$ and
$f_z(0)=0$, then $\circ$ is a sequential product on $({\cal
E}(H),0,I,\oplus)$, this result answered Gudder's problem
negatively.

\vskip 0.1 in

Now, we present a general method for constructing sequential
products on ${\cal E}(H)$.

\vskip 0.1 in

For each $A\in{\cal E}(H)$, take a $f_A\in B(sp(A))$.

Define $A\diamond B=f_A(A)B\overline{f_A}(A)$ for $A,B\in{\cal
E}(H)$.

\vskip 0.1 in

we say the set $\{f_A\}_{A\in{\cal E}(H)}$ satisfies {\it sequential
product condition} if the following two conditions hold:

(i) For every $A\in{\cal E}(H)$ and $t\in sp(A)$, $|f_A(t)|=
\sqrt{t}$;

(ii) For any $A,B\in{\cal E}(H)$, if $AB=BA$, then
$f_A(A)f_B(B)\approx f_{AB}(AB)$.

\vskip 0.1 in

If $\{f_A\}_{A\in{\cal E}(H)}$ satisfies sequential product
condition, then it is easy to see that

(1) $f_A(A)\overline{f_A}(A)=\overline{f_A}(A)f_A(A)=A$,
$(f_A(A))^*=\overline{f_A}(A)$.

(2) If $0\in sp(A)$, then $f_A(0)=0$.

(3) If $A=\sum\limits^{n}_{k=1}\lambda_{k}E_{k}$, where
$\{E_{k}\}^{n}_{k=1}$ are pairwise orthogonal projections, then
$f_A(A)=\sum\limits^{n}_{k=1}f_A(\lambda_{k})E_{k}$.

(4) For each $E\in {\cal P}(H)$,
$f_E(E)=f_E(0)(I-E)+f_E(1)E=f_E(1)E$.

(5) for any $A,B\in{\cal E}(H)$, $A\diamond B\in{\cal E}(H)$.

\vskip 0.1 in

{\bf Lemma 3.1 ([12])}. Let $H$ be a complex Hilbert space, $A,B\in
{\cal B}(H)$, $A,B,AB$ be three normal operators, and at least one
of $A,B$ be a compact operator. Then $BA$ is also a normal operator.

\vskip 0.1 in

{\bf Lemma 3.2 ([13])}. If $M,N,T\in {\cal B}(H)$, $M,N$ are normal
operators and $MT=TN$, then $M^*T=TN^*$.

\vskip 0.1 in

{\bf Lemma 3.3.} Suppose $\{f_A\}_{A\in{\cal E}(H)}$ satisfy
 sequential product condition and $A,B\in{\cal E}(H)$. If $A\diamond B=B\diamond A$
or $A\diamond B=\overline{f_B}(B)Af_B(B)$, then $AB=BA$.

{\bf Proof.} If $A\diamond B=B\diamond A$, that is,
$f_A(A)B\overline{f_A}(A)=f_B(B)A\overline{f_B}(B)$, then
$f_A(A)\overline{f_B}(B)f_B(B)\overline{f_A}(A)=f_B(B)\overline{f_A}(A)f_A(A)\overline{f_B}(B)$,
so $f_A(A)\overline{f_B}(B)$ is normal. By Lemma 3.1, we have
$\overline{f_B}(B)f_A(A)$ is also normal. Note that
$\Big{(}f_A(A)\overline{f_B}(B)\Big{)}f_A(A)=f_A(A)\Big{(}\overline{f_B}(B)f_A(A)\Big{)}$,
by using Lemma 3.2, we have
$\Big{(}f_A(A)\overline{f_B}(B)\Big{)}^*f_A(A)=f_A(A)\Big{(}\overline{f_B}(B)f_A(A)\Big{)}^*$.
That is, $f_B(B)A=Af_B(B)$. Taking adjoint, we have
$\overline{f_B}(B)A=A\overline{f_B}(B)$. Thus,
$AB=A\overline{f_B}(B)f_B(B)=\overline{f_B}(B)Af_B(B)=\overline{f_B}(B)f_B(B)A=BA$.

If $A\diamond B=\overline{f_B}(B)Af_B(B)$, that is,
$f_A(A)B\overline{f_A}(A)=\overline{f_B}(B)Af_B(B)$, the proof is
similar, we omit it.

\vskip 0.1 in

{\bf Lemma 3.4.} Suppose $\{f_A\}_{A\in{\cal E}(H)}$
 satisfy sequential product condition and $A,B\in{\cal E}(H)$. If
$AB=BA$, then $A\diamond B=B\diamond A=AB$.

{\bf Proof.} Since $AB=BA$, by sequential product condition (i) we
have $A\diamond B=f_A(A)B\overline{f_A}(A)=|f_A|^2(A)B=AB$.
Similarly, $B\diamond A=f_B(B)A\overline{f_B}(B)=|f_B|^2(B)A=AB$.
Thus $A\diamond B=B\diamond A=AB$.

\vskip 0.1 in

 {\bf Lemma 3.5.} Suppose $\{f_A\}_{A\in{\cal E}(H)}$
 satisfy sequential product condition and $A,B\in{\cal E}(H)$. If $AB=BA$, then for every $C\in
{\cal E}(H)$, $A\diamond (B\diamond C)=(A\diamond B)\diamond C$.

{\bf Proof.} By Lemma 3.4, $A\diamond B=AB$. By sequential product
condition (ii), there exists a complex constant $\xi$ such that
$|\xi| =1$ and $f_A(A)f_B(B)=\xi f_{AB}(AB)$. Taking adjoint, we
have $\overline{f_B}(B)\overline{f_A}(A)=\overline{\xi}\,
\overline{f_{AB}}(AB)$. Thus,
$f_A(A)f_B(B)C\overline{f_B}(B)\overline{f_A}(A)=f_{AB}(AB)C\overline{f_{AB}}(AB)=f_{A\diamond
B}(A\diamond B)C\overline{f_{A\diamond B}}(A\diamond B)$. That is,
$A\diamond (B\diamond C)=(A\diamond B)\diamond C$.

\vskip 0.1 in

{\bf Lemma 3.6 ([1]).} If $y,z\in H$ and $|\langle y,x\rangle
|=|\langle z,x\rangle |$ for every $x\in H$, then there exists a
$c\in {\mathbf{C}}$, $|c|=1$, such that $y=cz$.

\vskip 0.1 in

{\bf Lemma 3.7 ([14]).} Let $f: H\longrightarrow {\mathbf{C}}$ be a
mapping, $T\in {\cal B}(H)$. If the operator $S: H\longrightarrow H$
defined by $S(x)=f(x)T(x)$ is linear, then $f(x)=f(y)$ for every
$x,y\not\in Ker(T)$.

\vskip 0.1 in

{\bf Lemma 3.8.} Let $f: H\longrightarrow {\mathbf{C}}$ be a
mapping, $T\in {\cal B}(H)$. If the operator $S: H\longrightarrow H$
defined by $S(x)=f(x)T(x)$ is linear, then there exists a constant
$\xi\in{\mathbf{C}}$ such that $S(x)=\xi T(x)$ for every $x\in H$.

{\bf Proof.} By Lemma 3.7, there exists a constant
$\xi\in{\mathbf{C}}$ such that $S(x)=\xi T(x)$ for every $x\not\in
Ker(T)$. Of course, $S(x)=0=\xi T(x)$ for every $x\in Ker(T)$. So
$S(x)=\xi T(x)$ for every $x\in H$.

\vskip 0.1 in

Our main result in the section is the following.

\vskip 0.1 in

{\bf Theorem 3.1.} For each $A\in{\cal E}(H)$, take a $f_A\in
B(sp(A))$. Define $A\diamond B=f_A(A)B\overline{f_A}(A)$ for $A,
B\in{\cal E}(H)$. Then $\diamond$ is a sequential product on $({\cal
E}(H),0,I,\oplus)$ iff the set $\{f_A\}_{A\in{\cal E}(H)}$ satisfies
sequential product condition.

{\bf Proof.} (1) Firstly we suppose $\{f_A\}_{A\in{\cal E}(H)}$
satisfy sequential product condition, we show that $({\cal
E}(H),0,I,\oplus,\diamond)$ is a sequential effect algebra.

(SEA1) is obvious.

By Lemma 3.4, $I\diamond B=B$ for each $B\in{\cal E}(H)$, so (SEA2)
hold.

We verify (SEA3) as follows:

If $A\diamond B=0$, then $f_A(A)B\overline{f_A}(A)=0$, so
$f_A(A)B^{\frac{1}{2}}=0$, thus, we have
$AB=\overline{f_A}(A)f_A(A)B^{\frac{1}{2}}B^{\frac{1}{2}}=0$. Taking
adjoint, we have $AB=BA$. So $A\diamond B=B\diamond A$.

We verify (SEA4) as follows:

If $A\diamond B=B\diamond A$, then by Lemma 3.3, $AB=BA$. So
$A(I-B)=(I-B)A$. By Lemma 3.4, we have $A\diamond(I-B)=(I-B)\diamond
A$. By Lemma 3.5, $A\diamond (B\diamond C)=(A\diamond B)\diamond C$
for every $C\in {\cal E}(H)$.

We verify (SEA5) as follows:

If $C\diamond A=A\diamond C$ and $C\diamond B=B\diamond C$, then by
Lemma 3.3, $AC=CA$, $BC=CB$. So (SEA5) follows easily by Lemma 3.4.

Thus, we proved that $({\cal E}(H),0,I,\oplus,\diamond)$ is a
sequential effect algebra.

(2) Now we suppose $\diamond$ be a sequential product on $({\cal
E}(H),0,I,\oplus)$, we show that the set $\{f_A\}_{A\in{\cal E}(H)}$
satisfies sequential product condition.

Since $({\cal E}(H),0,I,\oplus,\diamond)$ is a sequential effect
algebra, by Theorem 2.2, for each $A\in {\cal E}(H)$, $A\diamond
I=A$, thus $|f_A|^2(A)=A$. If
$A=\sum\limits^{n}_{k=1}\lambda_{k}E_{k}$, where
$\{E_{k}\}^{n}_{k=1}$ are pairwise orthogonal projections,
$\sum\limits^{n}_{k=1}E_{k}=I$, then $sp(A)=\{\lambda_{k}\}$,
$|f_A|^2(A)=\sum\limits^{n}_{k=1}|f_A(\lambda_{k})|^2E_{k}$. Thus
$|f_A(\lambda_{k})|=\sqrt{\lambda_{k}}$ and $\{f_A\}_{A\in{\cal
E}(H)}$ satisfies sequential product condition (i).

To prove $\{f_A\}_{A\in{\cal E}(H)}$ satisfies sequential product
condition (ii), let $A,B\in {\cal E}(H)$ and $AB=BA$. By Theorem
2.2, we have $A\diamond B=B\diamond A=AB$. Thus by (SEA4),
$A\diamond (B\diamond C)=(A\diamond B)\diamond C$ for every $C\in
{\cal E}(H)$.

Let $x\in H$, $\|x\|=1$, $C=P_{x}$. Then for every $y\in H$, we have
$$\langle f_A(A)f_B(B)P_{x}\overline{f_B}(B)\overline{f_A}(A)y,y\rangle $$
$$=\langle \Big{(}A\diamond (B\diamond
P_{x})\Big{)}y,y\rangle $$
$$=\langle \Big{(}(A\diamond B)\diamond P_{x}\Big{)}y,y\rangle $$
$$=\langle \Big{(}(AB)\diamond P_{x}\Big{)}y,y\rangle $$
$$=\langle f_{AB}(AB)P_{x}\overline{f_{AB}}(AB)y,y\rangle  \ .$$

Since $$\langle
f_A(A)f_B(B)P_{x}\overline{f_B}(B)\overline{f_A}(A)y,y\rangle =
|\langle \overline{f_B}(B)\overline{f_A}(A)y,x\rangle |^2\ ,$$
$$\langle f_{AB}(AB)P_{x}\overline{f_{AB}}(AB)y,y\rangle =|\langle
\overline{f_{AB}}(AB)y,x\rangle |^2\ ,$$ we have $| \langle
\overline{f_B}(B)\overline{f_A}(A)y,x\rangle |=|\langle
\overline{f_{AB}}(AB)y,x\rangle |$ for every $x,y\in H$.

By Lemma 3.6, there exists a complex function g on $H$ such that
$|g(x)|\equiv 1$ and
$\overline{f_B}(B)\overline{f_A}(A)x=g(x)\overline{f_{AB}}(AB)x$ for
every $x\in H$. By Lemma 3.8, there exists a constant
$\xi\in{\mathbf{C}}$ such that $|\xi| =1$ and
$\overline{f_B}(B)\overline{f_A}(A)x=\xi\overline{f_{AB}}(AB)x$ for
every $x\in H$. So we conclude that
$\overline{f_B}(B)\overline{f_A}(A)=\xi\overline{f_{AB}}(AB)$.
Taking adjoint, we have $f_A(A)f_B(B)=\overline{\xi}f_{AB}(AB)$.
Thus $f_A(A)f_B(B)\approx f_{AB}(AB)$. This showed that the set
$\{f_A\}_{A\in{\cal E}(H)}$ satisfies sequential product condition.

\vskip 0.1 in

Theorem 3.1 present a general method for constructing sequential
products on ${\cal E}(H)$. Now, we give two examples.

\vskip 0.1 in

{\bf Example 3.1.} Let $g$ be a bounded complex Borel function on
$[0,1]$ such that
$$\hbox{$|g(t)|=\sqrt{t}$ for each $t\in [0,1]$\ ,}$$
$$\hbox{$g(t_1t_2)=g(t_1)g(t_2)$ for any $t_1,t_2\in [0,1]$\ .}$$

For each $A\in{\cal E}(H)$, let $f_A=g|_{sp(A)}$. Then it is easy to
know that $\{f_A\}$ satisfies sequential product condition. So by
Theorem 3.1, $A\diamond
B=f_A(A)B\overline{f_A}(A)=g(A)B\overline{g}(A)$ defines a
sequential product on the standard effect algebra $({\cal
E}(H),0,I,\oplus)$.

It is clear that Example 3.1 generalized Liu and Wu's result in [7].

\vskip 0.1 in

{\bf Example 3.2.} Let $H$ be a 2-dimensional complex Hilbert space,
$\Gamma=\{\gamma\mid \gamma$ be a decomposition of $I$ into two
rank-one orthogonal projections$\}$. For each $\gamma\in \Gamma$, we
can represent $\gamma$ by a pair of rank-one orthogonal projections
$(E_1,E_2)$, if $A\in{\cal E}(H)$, $A\not\in span\{I\}$ and
$A=\sum\limits^{2}_{k=1}\lambda_{k}E_{k}$, then we say that {\it $A$
can be diagonalized by $\gamma$}.

For each $\gamma\in \Gamma$, we take a $\xi(\gamma)\in \mathbf{R}$.
If $A\in{\cal E}(H)$, $A\not\in span\{I\}$ and $A$ can be
diagonalized by $\gamma$, let $f_A(t)=t^{\frac{1}{2}+\xi(\gamma)i}$
for $t\in sp(A)$.

If $A\in{\cal E}(H)$ and $A=\lambda I$, let $f_A(t)=\sqrt{t}$ for
$t\in sp(A)$.

Then the set $\{f_A\}_{A\in{\cal E}(H)}$ satisfies sequential
product condition (see the proof below). So by Theorem 3.1,
$A\diamond B=f_A(A)B\overline{f_A}(A)$ defines a sequential product
on the standard effect algebra $({\cal E}(H),0,I,\oplus)$.

{\bf Proof.} Obviously $\{f_A\}_{A\in{\cal E}(H)}$
  satisfies sequential product condition (i).

Now we show that $\{f_A\}_{A\in{\cal E}(H)}$ satisfies sequential
product condition (ii). Let $A,B\in{\cal E}(H)$, $AB=BA$.

(1) If $A=\sum\limits^{2}_{k=1}\lambda_{k}E_{k}$,
$B=\sum\limits^{2}_{k=1}\mu_{k}E_{k}$, $\lambda_{1}\neq\lambda_{2}$,
$\mu_{1}\neq\mu_{2}$, let $\gamma=(E_1,E_2)$, we have $f_A(t)=
t^{\frac{1}{2}+\xi(\gamma)i}$ for $t\in sp(A)$, $f_B(t)=
t^{\frac{1}{2}+\xi(\gamma)i}$ for $t\in sp(B)$. So
$f_A(A)=A^{\frac{1}{2}+\xi(\gamma)i}=\sum\limits^{2}_{k=1}\lambda_{k}^{\frac{1}{2}+\xi(\gamma)i}E_{k}$,
$f_B(B)=B^{\frac{1}{2}+\xi(\gamma)i}=\sum\limits^{2}_{k=1}\mu_{k}^{\frac{1}{2}+\xi(\gamma)i}E_{k}$.

(1a) If $\lambda_{1}\mu_{1}=\lambda_{2}\mu_{2}$, then
$AB=\lambda_{1}\mu_{1}I$, so $f_{AB}(t)=t^{\frac{1}{2}}$ for $t\in
sp(AB)$, thus we have
$f_{AB}(AB)=(AB)^{\frac{1}{2}}=\sqrt{\lambda_{1}\mu_{1}}I$,
$f_A(A)f_B(B)=\sum\limits^{2}_{k=1}(\lambda_{k}\mu_{k})^{\frac{1}{2}+\xi(\gamma)i}E_{k}
=(\lambda_{1}\mu_{1})^{\frac{1}{2}+\xi(\gamma)i}I=(\lambda_{1}\mu_{1})^{\xi(\gamma)i}f_{AB}(AB)\approx
f_{AB}(AB)$.

(1b) If $\lambda_{1}\mu_{1}\neq\lambda_{2}\mu_{2}$, then
$AB=\sum\limits^{2}_{k=1}\lambda_{k}\mu_{k}E_{k}$, so $f_{AB}(t)=
t^{\frac{1}{2}+\xi(\gamma)i}$ for $t\in sp(AB)$,
$f_{AB}(AB)=(AB)^{\frac{1}{2}+\xi(\gamma)i}=\sum\limits^{2}_{k=1}(\lambda_{k}\mu_{k})^{\frac{1}{2}+\xi(\gamma)i}E_{k}$,
thus we have
$f_A(A)f_B(B)=\sum\limits^{2}_{k=1}(\lambda_{k}\mu_{k})^{\frac{1}{2}+\xi(\gamma)i}E_{k}
=f_{AB}(AB)$.

(2) If $A=\lambda I$, $B=\sum\limits^{2}_{k=1}\mu_{k}E_{k}$,
$\mu_{1}\neq\mu_{2}$, let $\gamma=(E_1,E_2)$. Then we have
$f_A(t)=t^{\frac{1}{2}}$ for $t\in sp(A)$, $f_B(t)=
t^{\frac{1}{2}+\xi(\gamma)i}$ for $t\in sp(B)$. So
$f_A(A)=A^{\frac{1}{2}}=\sqrt{\lambda} I$,
$f_B(B)=B^{\frac{1}{2}+\xi(\gamma)i}=\sum\limits^{2}_{k=1}\mu_{k}^{\frac{1}{2}+\xi(\gamma)i}E_{k}$,
$AB=\sum\limits^{2}_{k=1}\lambda\mu_{k}E_{k}$.

(2a) If $\lambda=0$, then $AB=0$, $f_{AB}(t)=t^{\frac{1}{2}}$ for
$t\in sp(AB)$, so $f_{AB}(AB)=(AB)^{\frac{1}{2}}=0$. Thus
$f_A(A)f_B(B)=0=f_{AB}(AB)$.

(2b) If $\lambda\neq 0$, then
$f_{AB}(t)=t^{\frac{1}{2}+\xi(\gamma)i}$ for $t\in sp(AB)$. So
$f_{AB}(AB)=(AB)^{\frac{1}{2}+\xi(\gamma)i}=\lambda^{\frac{1}{2}+\xi(\gamma)i}\sum\limits^{2}_{k=1}(\mu_{k})^{\frac{1}{2}+\xi(\gamma)i}E_{k}$.
Thus
$f_A(A)f_B(B)=\sqrt{\lambda}\sum\limits^{2}_{k=1}(\mu_{k})^{\frac{1}{2}+\xi(\gamma)i}E_{k}
\approx f_{AB}(AB)$.

(3) If $A=\lambda I$, $B=\mu I$, then $f_A(t)=t^{\frac{1}{2}}$ for
$t\in sp(A)$, $f_B(t)=t^{\frac{1}{2}}$ for $t\in sp(B)$. So
$f_A(A)=A^{\frac{1}{2}}=\sqrt{\lambda} I$,
$f_B(B)=B^{\frac{1}{2}}=\sqrt{\mu} I$. $AB=\lambda\mu I$,
$f_{AB}(t)=t^{\frac{1}{2}}$ for $t\in sp(AB)$,
$f_{AB}(AB)=(AB)^{\frac{1}{2}}=\sqrt{\lambda\mu} I$. Thus
$f_A(A)f_B(B)=f_{AB}(AB)$.

It follows from (1)-(3) that the set $\{f_A\}_{A\in{\cal E}(H)}$
satisfies sequential product condition (ii).

\vskip 0.2 in

{\bf 4. Properties of the sequential product $\diamond$ on $({\cal
E}(H),0,I,\oplus)$}

\vskip 0.2 in

Now, we study some elementary properties of the sequential product
$\diamond$ defined in Section 3.

In this section, unless specified, we follow the notations in
Section 3. We always suppose $\{f_A\}_{A\in{\cal E}(H)}$ satisfies
sequential product condition. So by Theorem 3.1 $\diamond$ is a
sequential product on the standard effect algebra
$(\varepsilon(H),0,I,\oplus)$.

\vskip 0.1 in

{\bf Lemma 4.1.} If $C\in{\cal E}(H)$, $0\leq t\leq 1$, then
$f_{tC}(tC)\approx f_{tI}(t)f_{C}(C)$.

{\bf Proof.} Since $\{f_A\}_{A\in{\cal E}(H)}$
 satisfies sequential product condition, $f_{tC}(tC)\approx
f_{tI}(tI)f_{C}(C)=f_{tI}(t)f_{C}(C)$.

\vskip 0.1 in

{\bf Lemma 4.2.} Let $A\in{\cal E}(H)$, $x\in H$, $\|x\|=1$,
$\|f_A(A)x\|\neq 0$, $y=\frac{f_A(A)x}{\|f_A(A)x\|}$. Then
$A\diamond P_x=\|f_A(A)x\|^2P_y$.

{\bf Proof.} For each $z\in H$, $(A\diamond
P_x)z=f_A(A)P_x\overline{f_A}(A)z=\langle
\overline{f_A}(A)z,x\rangle f_A(A)x=\langle z,f_A(A)x\rangle
f_A(A)x=\|f_A(A)x\|^2P_yz$. So $A\diamond P_x=\|f_A(A)x\|^2P_y$.

\vskip 0.1 in

{\bf Lemma 4.3.} Let $M\subseteq {\cal B}(H)$ be a von Neumann
algebra, $P$ be a minimal projection in $M$, $A\in M$, $x\in Ran
(P)$, $\|x\|=1$. Then $PAP=\omega_x(A)P$, where $\omega_x(A)=\langle
Ax,x\rangle$.

{\bf Proof.} Since $P$ is a minimal projection in $M$, by [15,
Proposition 6.4.3], $PAP=\lambda P$ for some complex number
$\lambda$. Thus $\langle PAPx,x\rangle=\langle\lambda Px,x\rangle$.
So $\lambda=\omega_x(A)$.

\vskip 0.1 in

{\bf Theorem 4.1.} Let $A, B\in {\cal E}(H)$. Then the following
conditions are all equivalent:

(1) $AB=BA$;

(2) $A\diamond B=B\diamond A$;

(3) $A\diamond (B\diamond C)=(A\diamond B)\diamond C$ for every
$C\in {\cal E}(H)$.

{\bf Proof.} (1)$\Rightarrow$(2): By Theorem 2.2.

(2)$\Rightarrow$(1): By Lemma 3.3.

(1)$\Rightarrow$(3): By Lemma 3.5.

(3)$\Rightarrow$(1): Let $x\in H$, $\|x\|=1$, $C=P_{x}$. Then for
each $y\in H$,
$$\langle f_A(A)f_B(B)P_{x}\overline{f_B}(B)\overline{f_A}(A)y,y\rangle $$
$$=\langle \Big{(}A\diamond (B\diamond
P_{x})\Big{)}y,y\rangle $$
$$=\langle \Big{(}(A\diamond B)\diamond P_{x}\Big{)}y,y\rangle $$
$$=\langle f_{A\diamond B}(A\diamond B)P_{x}\overline{f_{A\diamond B}}(A\diamond B)y,y\rangle \ .$$

Since $$\langle
f_A(A)f_B(B)P_{x}\overline{f_B}(B)\overline{f_A}(A)y,y\rangle =|
\langle \overline{f_B}(B)\overline{f_A}(A)y,x\rangle | ^2\ ,$$
$$\langle f_{A\diamond B}(A\diamond B)P_{x}\overline{f_{A\diamond B}}(A\diamond B)y,y\rangle =| \langle
\overline{f_{A\diamond B}}(A\diamond B)y,x\rangle | ^2\ ,$$ we have
$| \langle\overline{f_B}(B)\overline{f_A}(A)y,x\rangle | =| \langle
\overline{f_{A\diamond B}}(A\diamond B)y,x\rangle |$ for every
$x,y\in H$.

\vskip 0.1 in

By Lemma 3.6, there exists a complex function $g$ on $H$ such that
$|g(x)|=1$ and
$\overline{f_B}(B)\overline{f_A}(A)x=g(x)\overline{f_{A\diamond
B}}(A\diamond B)x$ for every $x\in H$.

By Lemma 3.8, there exists a constant $\xi$ such that $| \xi| =1$
and $\overline{f_B}(B)\overline{f_A}(A)x=\xi\overline{f_{A\diamond
B}}(A\diamond B)x$ for every $x\in H$.

\vskip 0.1 in

So we conclude that
$\overline{f_B}(B)\overline{f_A}(A)=\xi\overline{f_{A\diamond
B}}(A\diamond B)$.

Taking adjoint, we have $f_A(A)f_B(B)=\overline{\xi}f_{A\diamond
B}(A\diamond B)$. Thus
$\overline{f_B}(B)Af_B(B)=\overline{f_B}(B)\overline{f_A}(A)f_A(A)f_B(B)=\xi\overline{f_{A\diamond
B}}(A\diamond B)\overline{\xi}f_{A\diamond B}(A\diamond B)=A\diamond
B$. That is, $A\diamond B=\overline{f_B}(B)Af_B(B)$, so by Lemma
3.3, we have $AB=BA$.

\vskip 0.1 in

{\bf Theorem 4.2.} Let $A, B\in {\cal E}(H)$. If $A\diamond B\in
{\cal P}(H)$, then $AB=BA$.

{\bf Proof.} If $A\diamond B=0$, then by (SEA3) we have $A\diamond
B=B\diamond A$, so by Theorem 4.1 we have $AB=BA$.

If $A\diamond B\neq 0$. Firstly, we let $x\in Ran(A\diamond B)$ and
$\|x\|=1$. Then $f_A(A)B\overline{f_A}(A)x=x$. So $\langle
B\overline{f_A}(A)x,\overline{f_A}(A)x\rangle =1$. By Schwarz
inequality, we conclude that
$B\overline{f_A}(A)x=\overline{f_A}(A)x$. Thus
$Ax=f_A(A)\overline{f_A}(A)x=f_A(A)B\overline{f_A}(A)x=x$. So $1\in
sp(A)$ and
$B\overline{f_A}(A)x=\overline{f_A}(A)x=\overline{f_A}(1)x$.

Next, we let $x\in Ker(A\diamond B)$ and $\|x\|=1$. Then
$f_A(A)B\overline{f_A}(A)x=0$. So $\langle
B\overline{f_A}(A)x,\overline{f_A}(A)x\rangle =0$. We conclude that
$B\overline{f_A}(A)x=0$.

Thus, we always have $B\overline{f_A}(A)=\overline{f_A}(1)(A\diamond
B)$. That is, $f_A(1)B\overline{f_A}(A)=A\diamond B$.

Taking adjoint, we have
$f_A(1)B\overline{f_A}(A)=\overline{f_A}(1)f_A(A)B$.

By Lemma 3.2, we have
$\overline{f_A}(1)Bf_A(A)=f_A(1)\overline{f_A}(A)B$. So
$f_A(1)\overline{f_A}(A)B$ is self-adjoint. By [15, Proposition
3.2.8], we have

$sp\Big{(}f_A(1)\overline{f_A}(A)B\Big{)}\backslash\{0\}=sp\Big{(}f_A(1)B\overline{f_A}(A)\Big{)}\backslash\{0\}=sp(A\diamond
B)\backslash\{0\}\subseteq \mathbf{R^+}$.

Thus we conclude that $f_A(1)\overline{f_A}(A)B\geq 0$.

\vskip 0.1 in

Since
$\Big{(}f_A(1)\overline{f_A}(A)B\Big{)}^2=\Big{(}\overline{f_A}(1)Bf_A(A)\Big{)}\Big{(}
f_A(1)\overline{f_A}(A)B\Big{)}=BAB=\Big{(}f_A(1)B\overline{f_A}(A)\Big{)}\Big{(}\overline{f_A}(1)f_A(A)B\Big{)}=(A\diamond
B)^2$, by the uniqueness of positive square root, we have
$f_A(1)\overline{f_A}(A)B=A\diamond B$. That is,
$f_A(1)\overline{f_A}(A)B=\overline{f_A}(1)Bf_A(A)=f_A(1)B\overline{f_A}(A)=\overline{f_A}(1)f_A(A)B=A\diamond
 B$. Thus,
$BA=f_A(1)B\overline{f_A}(A)\overline{f_A}(1)f_A(A)=f_A(1)\overline{f_A}(A)B\overline{f_A}(1)f_A(A)=f_A(1)\overline{f_A}(A)\overline{f_A}(1)f_A(A)B=AB$.

\vskip 0.1 in

{\bf Theorem 4.3.} Let $A, B\in {\cal E}(H)$. Then the following
conditions are all equivalent:

(1) $A\diamond(C\diamond B)=(A\diamond C)\diamond B$ for every $C\in
{\cal E}(H)$;

(2) $C\diamond(A\diamond B)=(C\diamond A)\diamond B$ for every $C\in
{\cal E}(H)$;

(3) $\langle (A\diamond B)x,x\rangle =\langle Ax,x\rangle \langle
Bx,x\rangle $ for every $x\in H$ with $\|x\|=1$;

(4) $A=tI$ or $B=tI$ for some $0\leq t\leq 1$.

{\bf Proof.} By Theorem 2.4, we conclude that
(2)$\Longleftrightarrow$(3)$\Longleftrightarrow$(4).

(4)$\Rightarrow$(1) follows from Lemma 2.4 and Theorem 2.2 easily.

(1)$\Rightarrow$(4): If (1) hold, then $A\diamond (P_x\diamond
B)=(A\diamond P_x)\diamond B$ for each $x\in H$ with $\|x\|=1$.
Without lose of generality, we suppose $\|f_A(A)x\|\neq 0$. Let
$y=\frac{f_A(A)x}{\|f_A(A)x\|}$.

By Lemma 4.2 and Theorem 2.1, $$A\diamond (P_x\diamond
B)=f_A(A)(P_xBP_x)\overline{f_A}(A)$$
$$=f_A(A)(\langle Bx,x\rangle
P_x)\overline{f_A}(A)$$
$$=\langle Bx,x\rangle (A\diamond P_x)$$
$$=
\|f_A(A)x\|^2\langle Bx,x\rangle P_y\ .$$

By Lemma 4.1 and Lemma 4.2,
$$(A\diamond P_x)\diamond B=(\|f_A(A)x\|^2P_y)\diamond
B$$
$$=f_{\|f_A(A)x\|^2P_y}(\|f_A(A)x\|^2P_y)B\overline{f_{\|f_A(A)x\|^2P_y}}(\|f_A(A)x\|^2P_y)$$
$$=f_{\|f_A(A)x\|^2I}(\|f_A(A)x\|^2)f_{P_y}(P_y)B\overline{f_{\|f_A(A)x\|^2I}}(\|f_A(A)x\|^2)\overline{f_{P_y}}(P_y)$$
$$=\|f_A(A)x\|^2P_yBP_y$$
$$=\|f_A(A)x\|^2\langle
By,y\rangle P_y\ .$$ Thus $\langle Bx,x\rangle =\langle By,y\rangle
$. So we have $\langle \overline{f_A}(A)Bf_A(A)x,x\rangle =\langle
Ax,x\rangle \langle Bx,x\rangle $. By Lemma 2.5, we conclude that
(4) hold.

\vskip 0.1 in

{\bf Theorem 4.4.} Let $A\in {\cal E}(H)$, $E\in {\cal P}(H)$. Then
the following conditions are all equivalent:

(1) $A\diamond E\leq E$;

(2) $E\overline{f_A}(A)(I-E)=0$.

{\bf Proof.} Since $E\in {\cal P}(H)$ and $\|\overline{f_A}(A)\|\leq
1$, we have
$$A\diamond E\leq E\Longleftrightarrow \langle
f_A(A)E\overline{f_A}(A)x,x\rangle\leq\langle Ex,x\rangle \hbox{ for
every $x\in H$}$$
$$\Longleftrightarrow
\|E\overline{f_A}(A)x\|\leq\|Ex\| \hbox{ for every $x\in H$}$$
$$\Longleftrightarrow
\overline{f_A}(A)\mid_{Ker(E)}\subseteq Ker(E)$$
$$\Longleftrightarrow
E\overline{f_A}(A)(I-E)=0\ .$$

\vskip 0.1 in

{\bf Corollary 4.1 [14].} Let $A\in {\cal E}(H)$, $E\in {\cal
P}(H)$. Then the following conditions are all equivalent:

(1) $A^{\frac{1}{2}}EA^{\frac{1}{2}}\leq E$;

(2) $AE=EA$.

{\bf Proof.} (2)$\Rightarrow$(1) is trivial.

(1)$\Rightarrow$(2): Let $f_B(t)=\sqrt{t}$ for each $B\in {\cal
E}(H)$ and $t\in sp(B)$, then $\{f_B\}_{B\in {\cal E}(H)}$ satisfies
sequential product condition. For this sequential product,
$A\diamond E=A^{\frac{1}{2}}EA^{\frac{1}{2}}$. So by Theorem 4.4 we
have $EA^{\frac{1}{2}}(I-E)=0$. That is,
$EA^{\frac{1}{2}}=EA^{\frac{1}{2}}E$. Taking adjoint, we have
$EA^{\frac{1}{2}}=A^{\frac{1}{2}}E$. Thus $AE=EA$.

\vskip 0.1 in

{\bf Corollary 4.2.} Let $M\subseteq {\cal B}(H)$ be a von Neumann
algebra, ${\cal E}(M)=\{A\in M|0\leq A\leq I\}$, $P$ or $I-P$ be a
minimal projection in $M$. Then for every $A\in{\cal E}(M)$, the
following conditions are all equivalent:

(1) $A\diamond P\leq P$;

(2) $AP=PA$.

{\bf Proof.} (2)$\Rightarrow$(1): By Theorem 2.2, $A\diamond
P=AP=PAP\leq P$.

(1)$\Rightarrow$(2): If $P$ is a minimal projection in $M$, then by
Theorem 4.4 we have $P\overline{f_A}(A)(I-P)=0$, that is,
$P\overline{f_A}(A)=P\overline{f_A}(A)P$.

Let $x\in Ran (P)$ with $\|x\|=1$. Then by Lemma 4.3 we have
$P\overline{f_A}(A)P=\omega_x(\overline{f_A}(A))P$. So
$P\overline{f_A}(A)=\omega_x(\overline{f_A}(A))P$. Taking adjoint,
we have $f_A(A)P=\omega_x(f_A(A))P$. By Lemma 3.2, we have
$Pf_A(A)=\overline{\omega_x(\overline{f_A}(A))}P=\omega_x(f_A(A))P$.
Thus $Pf_A(A)=f_A(A)P$. Taking adjoint, we have
$P\overline{f_A}(A)=\overline{f_A}(A)P$. So,
$PA=Pf_A(A)\overline{f_A}(A)=f_A(A)P\overline{f_A}(A)=f_A(A)\overline{f_A}(A)P=AP$.

\vskip 0.1 in

If $I-P$ is a minimal projection in $M$. By Theorem 4.4 we have
$P\overline{f_A}(A)(I-P)=0$. Taking adjoint, we have
$(I-P)f_A(A)P=0$. That is, $(I-P)f_A(A)=(I-P)f_A(A)(I-P)$. Similar
to the proof above, we conclude that $(I-P)A=A(I-P)$. So $AP=PA$.

\vskip0.2in

\centerline{\bf References}

\vskip0.2in

\noindent [1]. Gudder, S, Nagy, G. Sequential quantum measurements.
J. Math. Phys. 42(2001), 5212-5222.

\noindent [2]. Gudder, S, Greechie, R. Sequential products on effect
algebras. Rep. Math. Phys.  49(2002), 87-111.

\noindent [3]. Gheondea, A, Gudder, S. Sequential product of quantum
effects. Proc. Amer. Math. Soc. 132 (2004), 503-512.

\noindent [4]. Gudder, S. Open problems for sequential effect
algebras. Inter. J. Theory. Phys. 44 (2005), 2219-2230.

\noindent [5]. Gudder, S, Latr¨¦moli¨¨re, F. Characterization of the
sequential product on quantum effects. J. Math. Phys. 49 (2008),
052106-052112.

\noindent [6]. Shen J., Wu J. D.. Not each sequential effect algebra
is sharply dominating. Phys. Lett. A 373 (2009), 1708-1712.

\noindent [7]. Liu Weihua, Wu Junde. A uniqueness problem of the
sequence product on operator effect algebra ${\cal E} (H)$. J. Phys.
A: Math. Theor. 42(2009), 185206-185215.

\noindent [8]. Foulis, D J, Bennett, M K. Effect algebras and
unsharp quantum logics. Found. Phys. 24 (1994), 1331-1352.

\noindent [9]. Gudder, S. Sharply dominating effect algebras. Tatra
Mt. Math. Publ. 15(1998), 23-30.

\noindent [10]. Riecanova, Z, Wu Junde. States on sharply dominating
effect algebras. Science in China Series A: Math. 51(2008), 907-914.

\noindent [11]. Smuljan, J L. An operator Hellinger integral
(Russian). Mat. Sb. (N.S.) 49(1959), 381-430.

\noindent [12]. Kaplansky, I. Products of normal operators. Duke
Math. J. 20 (1953), 257-260.

\noindent [13]. Rudin, W. Functional analysis. McGraw-Hill, New York
(1991).

\noindent [14]. Li, Y, Sun, X H, Chen, Z L. Generalized infimum and
sequential product of quantum effects. J. Math. Phys. 48 (2007),
102101.

\noindent [15]. Kadison, R, Ringrose, J. Fundamentals of the theory
of operator algebras (I, II). American Mathematical Society, New
York (1997).

\end{document}